\documentclass[superscriptaddress]{revtex4}

\usepackage{amsmath}
\usepackage{amssymb}
\usepackage[latin9]{inputenc}
\usepackage{graphicx}
\usepackage{color}
\usepackage{MnSymbol}

\def \be {\begin{equation}}
\def \ee {\end{equation}}
\def \bea {\begin{eqnarray}}
\def \eea {\end{eqnarray}}

\makeatletter \@ifundefined{textcolor}{} {
 \definecolor{BLACK}{gray}{0}
 \definecolor{WHITE}{gray}{1}
 \definecolor{RED}{rgb}{1,0,0}
 \definecolor{GREEN}{rgb}{0,1,0}
 \definecolor{BLUE}{rgb}{0,0,1}
 \definecolor{CYAN}{cmyk}{1,0,0,0}
 \definecolor{MAGENTA}{cmyk}{0,1,0,0}
 \definecolor{YELLOW}{cmyk}{0,0,1,0}
 \definecolor{GOLDEN}{rgb}{0.85,.66,0}
 }
\makeatother

\begin{document}

\title{High precision solutions to quantized vortices within Gross-Pitaevskii equation}

\author{Hao-Hao Peng}
\email{penghh@mail.ustc.edu.cn}
\affiliation{Department of Modern Physics, University of Science and Technology of China, Hefei, Anhui 230026, China}

\author{Jian Deng}
\email{jdeng@sdu.edu.cn}
\affiliation{Institute of Frontier and Interdisciplinary Science,
Key Laboratory of Particle Physics and Particle Irradiation (MOE), Shandong University,
Qingdao, Shandong 266237, China}

\author{Sen-Yue Lou}
\email{lousenyue@nbu.edu.cn}
\affiliation{School of Physical Science and Technology, Ningbo University, Ningbo, 315211, China}

\author{Qun Wang}
\email{qunwang@ustc.edu.cn}
\affiliation{Department of Modern Physics, University of Science and Technology of China, Hefei, Anhui 230026, China}

\begin{abstract}
The dynamics of vortices in Bose-Einstein condensates of dilute cold atoms can be well formulated by Gross-Pitaevskii equation. To better understand the properties of vortices, a systematic method to solve the nonlinear differential equation for the vortex to a very high precision is proposed. Through two-point Pad$\acute{\text{e}}$ approximants, these solutions are presented in terms of simple rational functions, which can be used in the simulation of vortex dynamics. The precision of the solutions is sensitive to the connecting parameter and the truncation orders. It can be improved significantly with a reasonable extension in the order of rational functions. The errors of the solutions and the limitation of two-point Pad$\acute{\text{e}}$ approximants are discussed. This investigation may shed light on the exact solution to the nonlinear vortex equation.
\end{abstract}


\maketitle

\section{Introduction}
Vortices are strongly nonlinear excitations in superfluid, which are quantized as topologic defects from the long range quantum phase coherence \cite{BOOK:2003,Fetter:2009zz,BOOK:2016}. The formation, stability and dynamical properties of vortices have been intensively studied experimentally \cite{Matthews:1999zz, PhysRevLett.89.190403,Inouye:2001, Shaeer:2001,Scherer:2007} and theoretically \cite{Carretero:2008, Deng:2022wyz}. Vortices play important roles not only in many-body quantum systems, but also in dark matter \cite{Berezhiani:2015pia,Berezhiani:2015bqa,Hui:2020hbq} and the phase transition of the early universe \cite{Kibble:1976sj,Zurek:1985qw,Hindmarsh:1994re,Zurek:1996sj}. The structure of nonlinear dynamics for vortices is in connection with the gravitational field equation for the metric of black holes, which makes it possible to study the Penrose process of rotating black holes in laboratories \cite{Garay:1999sk,Solnyshkov:2018dgq}. Vortices are also necessary degrees of freedom in turbulence and the intermediate state of an over-populated off-equilibrium system. The generation and clustering of vortices and the annihilation of vortex-antivortex pairs are closely related to the scaling law of a non-thermal fixed point \cite{Berges:2008wm,Berges:2008sr,Berges:2014bba,Nowak:2011sk,Schole:2012kt,Karl:2016wko,Deng:2018xsk}.

The static and dynamical properties of vortices and vortex lattices at zero temperature limit can be well described by the Gross-Pitaevskii equation (GPE), which is a nonlinear Schrödinger equation with the nonlinearity being determined by interactions through a mean-field approximation \cite{Gross:1963,Pitaevkii:1961}. This equation emerges in various nonlinear phenomena, and has been extensively studied in connection with nonlinear optics, plasma physics and fluid dynamics. The GPE helps the prediction and description on experimental observations of nonlinear effects, such as vortices and interaction among them \cite{BOOK:Emergent2008,BOOK:NLSE2019}.   

As an individual object, the vortex has an internal structure and can also be involved in an external evolution. How the wave function of a vortex responds to a perturbation or to the nonuniformity of the condensate is an interesting and complicated problem. The structure of the vortex core is determined by  the properties of the vortex itself, but it should also match the outer wave function determined by the collective motion and interaction with the environment. Through a unique coordinate transform and matching the asymptotic expansion inside and outside the vortex core, a set of differential equations have been derived to describe the overall structure and collective motion of the vortex in a perturbative way \cite{Svidzinsky:2000,Koens:2012}.

To construct a perturbation theory, the ideal starting point is a system with an exact solution. If the disturbance is not large, a series of perturbative solutions can be obtained systematically, so that a reliable understanding of the complex system can be achieved. The various physical quantities associated with the perturbed system can be expressed as corrections to the original one. For example, the contribution from the trace anomaly of the QED energy momentum tensor to the mass of the hydrogen atom, the simplest QED bound state, is calculated in Ref. \cite{Sun:2020ksc}, where the Lamb shift from the vacuum polarization and radiative corrections to the electron mass are perturbative corrections to the ground state with an exact wave function.

To study the vortex dynamics in a perturbative way, the stationary wave function of a vortex in a uniform condensate is the starting point. But due to the nonlinear nature of this problem, an exact solution to the nonlinear ordinary differential equations (ODE) has not been found yet. The best alternative we can think of is to look for an approximate function that is simple, compact and as precise as possible. This can be done with the method of two-point Pad$\acute{\text{e}}$ approximants. The basic idea of Pad$\acute{\text{e}}$ approximants is to construct a rational function which can reproduce the Taylor expansion series of the target function within a given order. The rational function can mimic the singularity of the target function since it contains pole structure, so the convergence radius of the original Taylor expansion can be extended (see an example in \cite{Tian:2021}). If the target function is constrained by two boundary conditions, the rational function should reproduce two Taylor expansion series simultaneously, which is called two-point Pad$\acute{\text{e}}$ approximants \cite{Mccabe:1976,Sidi:1980}. We found that this scheme is particularly suitable for finding approximate solutions to the nonlinear equation for the vortex. The main purpose of this paper is to explore the precision limit of this scheme, hoping to provide a clue for the exact solution of the vortex.


The outline of this paper is as follows.  In Sec. \ref{Sec:vortex-GPE}, the static and rotationally symmetric equation for the voetex is derived. In Sec.\ref{Sec:relaxation}, a very high precision numerical solution to the vortex equation is obtained with a relaxation method, which can be used to test the accuracy of the approximate solutions. In Sec. \ref{Sec:2-point-pade}, the results from two-point Pad$\acute{\text{e}}$ approximants are presented. In Sec. \ref{Sec:error}, the errors and challenge of this approach are discussed.  The conclusions and discussions are given in Sec. \ref{Sec:summary}.


\section{Vortices within Gross-Pitaevskii Equation}
\label{Sec:vortex-GPE}
The Bose-Einstein condensate (BEC) in cold atom systems at zero temperature can be well described by the Gross-Pitaevskii(GP) theory as the mean-filed approximation of quantum field theories \cite{BOOK:2003}. In GP theory, the ground state and weakly excited states of the condensate are described by the complex wave function $\psi(\mathbf{r},t)$, which satisfies GPE,
\bea
\text{i} \hbar \frac{\partial}{\partial t}\psi(\mathbf{r},t) = -\frac{\hbar^2}{2m}\nabla^2 \psi(\mathbf{r},t) + \lambda |\psi(\mathbf{r},t)|^2 \psi(\mathbf{r},t),
\eea
where $\hbar$ is the reduced Planck constant, $m$ is the particle mass, and the coupling constant $\lambda$ can reproduce the $s$-wave scattering length, $\lambda= 4\pi a_s \hbar^2/m$ in the Born approximation. The investigation on the exact and numerical solutions to GPE improves our understanding on the nonlinear dynamics of matter waves in BEC \cite{BOOK:NLSE2019}. In this work, we will focus on the static and rotationally symmetric solution to GPE, the quantum vortex, a kind of topological excitation in a superfluid in presence of the local orbital angular momentum. The wave function can be written in the following form
\bea
\psi(\mathbf{r},t) = \sqrt{n_0} f\left( \eta = \frac{|\mathbf{r}|}{\xi}\right) \text{e}^{\text{i} s \varphi} \text{e}^{-\text{i} \mu t},
\eea
where $n_0$ is the bulk number density, $\xi = 1/\sqrt{2m \lambda n_0}$ is the healing length, $\varphi$ is the azimuthal angle, $\mu$ is the chemical potential, $s$ is the winding number which must be an integer to keep the wave function single-valued. The normalized amplitude function $f(\eta)$ satisfies the nonlinear ordinary differential equation (ODE)
\bea
\frac{1}{\eta} \frac{\text{d}}{\text{d}\eta} \left[ \eta\frac{\text{d} f(\eta)}{\text{d}\eta} \right]
+ \left( \frac{\mu}{\lambda n_0}-\frac{s^2}{\eta^2}\right) f(\eta) - f(\eta)^3 =0.
\eea
Physically, the area modified by a vortex is rather limited, the density profile should return to the bulk value in a region far away from the vortex core, which requires $\mu/\lambda n_0=1$. So the ODE can be put into the form
\begin{equation}
f''(\eta)+\frac{1}{\eta} f'(\eta) + (1-\frac{s^2}{\eta^2})f(\eta) - f(\eta)^3 =0,  \label{eq:vortex}
\end{equation}
with the boundary conditions $f(\eta \rightarrow 0)=k_s \eta^s$ and $f(\eta\rightarrow\infty)=1$. Here $k_s$ is the connecting parameter, which plays an important role in the vortex solution.


The purpose of this work is to solve Eq. \eqref{eq:vortex}. Without the cubic term, the ODE becomes the Bessel's differential equation with Bessel functions as the formal solution, but the boundary condition at infinity cannot be satisfied. With the cubic term, however, the equation is a nonlinear ODE. It is very hard, if not impossible, to obtain the exact analytical solution to it. So far, we can only find a similar example with an exact solution. It can be easily checked that $f=\tanh(\eta/\sqrt{2})$ is one solution to the nonlinear ODE, $f'' +f(1 - f^2) =0$ with the same boundary conditions. This solution is a direct extension of the Riccati equation. In Appendix \ref{Sec:tanh}, we will demonstrate another way to obtain this exact solution which may provide some hints for solving Eq. \eqref{eq:vortex}. The example gives us a little hope that a sufficiently precise numerical solution can guide the guesswork about the possible form of the analytical solution.

\begin{figure}[ht]
\centering
\includegraphics[scale=0.4]{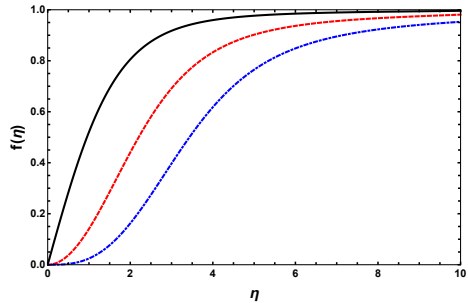}
\caption{The profile function $f(\eta)$ are obtained with shooting method in the range $\eta< \eta_C = 10$. Black solid line, red dashed line and blue dash-dotted line stand for the results with $s=1,2,3$ respectively.}
\label{fig:vortex-GPE}
\end{figure}

For the numerical solution, the differential equation with two-point boundary conditions can be solved with shooting method, which requires a fine tuning of $k_s$ to achieve the right limit $f(\eta \rightarrow \eta_C)\rightarrow 1$ and $f'(\eta \rightarrow \eta_C)\rightarrow 0$ in the range $\eta<\eta_C$. The shooting results of $f(\eta)$ for the winding numbers $s=1,2,3$ are shown in FIG. \ref{fig:vortex-GPE}. It seems that shooting works well, but the results are not stable for a large $\eta_C$, because a rigorous precision in $k_s$ and an inaccessible small step size in the finite difference are required. Fortunately, we find a relaxation method which can achieve a very high precision in the whole region of $0< \eta < \infty$. The technical details will be discussed in the next section.

\section{Relaxation iteration for a high precision numerical solution}
\label{Sec:relaxation}
A high precision numerical solution to Eq. \eqref{eq:vortex} can be achieved by a relaxation iteration. The method is built up with three steps. Firstly, we take a replacement for the target function, $f(\eta)  \rightarrow  \eta^s/(\eta^s+1) + h(\eta)$. The auxiliary function $\eta^s/(\eta^s+1)$ is to make the unknown function $h(\eta )$ satisfy the boundary condition $h(0)=h(\infty)=0$, which is convenient for solving the ODE. The function $h(\eta)$ satisfies the nonlinear ODE as follows
\bea
\frac{\text{d}^2h}{\text{d}\eta^2} + \frac{1}{\eta}\frac{\text{d}h}{\text{d}\eta} + \left[1-\frac{s^2}{\eta^2}-\frac{3\eta^{2s}}{(1+\eta^s)^2}\right] h = h^3 + \frac{3\eta^s}{1+\eta^s} h^2 +\frac{\eta^s\left(s^2 \eta^s (\eta^s+3) -\eta^2(2\eta^s+1)\right)}{\eta^2(1+\eta^s)^3},  \label{Eq-g0}
\eea
where we have put all linear terms of $h(x)$ to the left-hand-side.
Secondly, we replace the variable to $x=\eta /(R+\eta )$, so that the limit $\eta \rightarrow \infty$ is shifted to $x\rightarrow 1$. The positive number $R$ sets a finite scale, we choose $R=1$ without loss of generality.
The nontrivial change is
\bea
\frac{\text{d}^2h}{\text{d}\eta^2} + \frac{1}{\eta}\frac{\text{d}h}{\text{d}\eta} \rightarrow \frac{(1-x)^4}{R^2}\frac{\text{d}^2h}{\text{d}x^2} -2\frac{(1-x)^3}{R^2}\frac{\text{d}h}{\text{d}x} +\frac{(1-x)^3}{R^2 x}\frac{\text{d}h}{\text{d}x}.
\eea
In this way, the ODE is ready for the discretization and iteration.


Finally we discretize $x$ as $x_i=i \Delta_x$, where $i=0,\cdots, N$ and the step size is $\Delta_x =1/N$. So the derivatives become
\bea
\frac{\text{d}^2h}{\text{d}x^2}&=&\frac{h_{i+1}+h_{i-1}-2h_i}{\Delta_x^2},\nonumber\\
\frac{\text{d}h}{\text{d}x}&=& \frac{h_{i+1}-h_{i-1}}{2 \Delta_x}.
\eea
Then the nonlinear ODE is changed to a set of linear algebraic equation
\be
a_i h_{i-1} +b_i h_i + c_i h_{i+1} = w_i(h_i), \label{disc-linear}
\ee
for $i=1,\cdots, N-1$, where $a_i, b_i, c_i, w_i(h_i)$ are well defined at each point.
Because of the well-designed boundary condition, $h_0=h_{N}=0$ doesn't interfere the iteration, then Eq. (\ref{disc-linear}) can be rewritten as
\bea
\left(
\begin{array}{ccc cccc}
 b_1 & c_1 & 0   & \cdots & 0 & 0 & 0 \\
 a_2 & b_2 & c_2 & \cdots & 0 & 0 & 0 \\
 0   & a_2 & b_3 & \cdots & 0 & 0 & 0 \\
 \vdots   & \vdots   & \vdots & \ddots & \vdots   & \vdots   & \vdots \\
 0   & 0 & 0 & \cdots  & b_{N-3} & c_{N-3} & 0 \\
 0   & 0 & 0 & \cdots  & a_{N-2} & b_{N-2} & c_{N-2} \\
 0   & 0 & 0 & \cdots  & 0 & a_{N-1} & b_{N-1}  \\
\end{array}
\right)
\left(
\begin{array}{c}
h_1\\
h_2\\
h_3\\
\vdots \\
h_{N-3} \\
h_{N-2} \\
h_{N-1} \\
\end{array}
\right)
=
\left(
\begin{array}{c}
w_1\\
w_2\\
w_3\\
\vdots \\
w_{N-3} \\
w_{N-2} \\
w_{N-1} \\
\end{array}
\right). \label{eq:matrix-form}
\eea
We may start by choosing a set of reasonable values for $h_i$ and obtain $w_i(h_i)$, then a new set of $h_i$ can be determined by solving the tridiagonal algebraic equation \eqref{eq:matrix-form}. In this way, a high precision solution can be achieved within dozens of iterations. In the last iteration, the maximum error to $h_i$ is smaller than $10^{-25}$, which means that the algorithm has a fast speed of convergence. The main source to the error comes from the discretization, which can be improved by increasing the number of points $N$. We find the maximum difference between the results with $N=2^{25}$ and $N=2^{26}$ is of the order $10^{-16}$, which sets the precision of our numerical solution. The main challenge to go beyond larger $N$ is to handle a larger dimension vector with high precision on a personal computer, that is the reason we stopped at $N=2^{26}$. For the vector with such a dimension, the iteration to achieve the required precision can be done in a few hours. 


For applications, the numerical solution in the form of data sample is not convenient.
It is better to look for an approximate analytical solution. The present numerical solution can be used to verify the accuracy of the approximate analytical solution.

\begin{figure}[ht]
\centering
\includegraphics[scale=0.40]{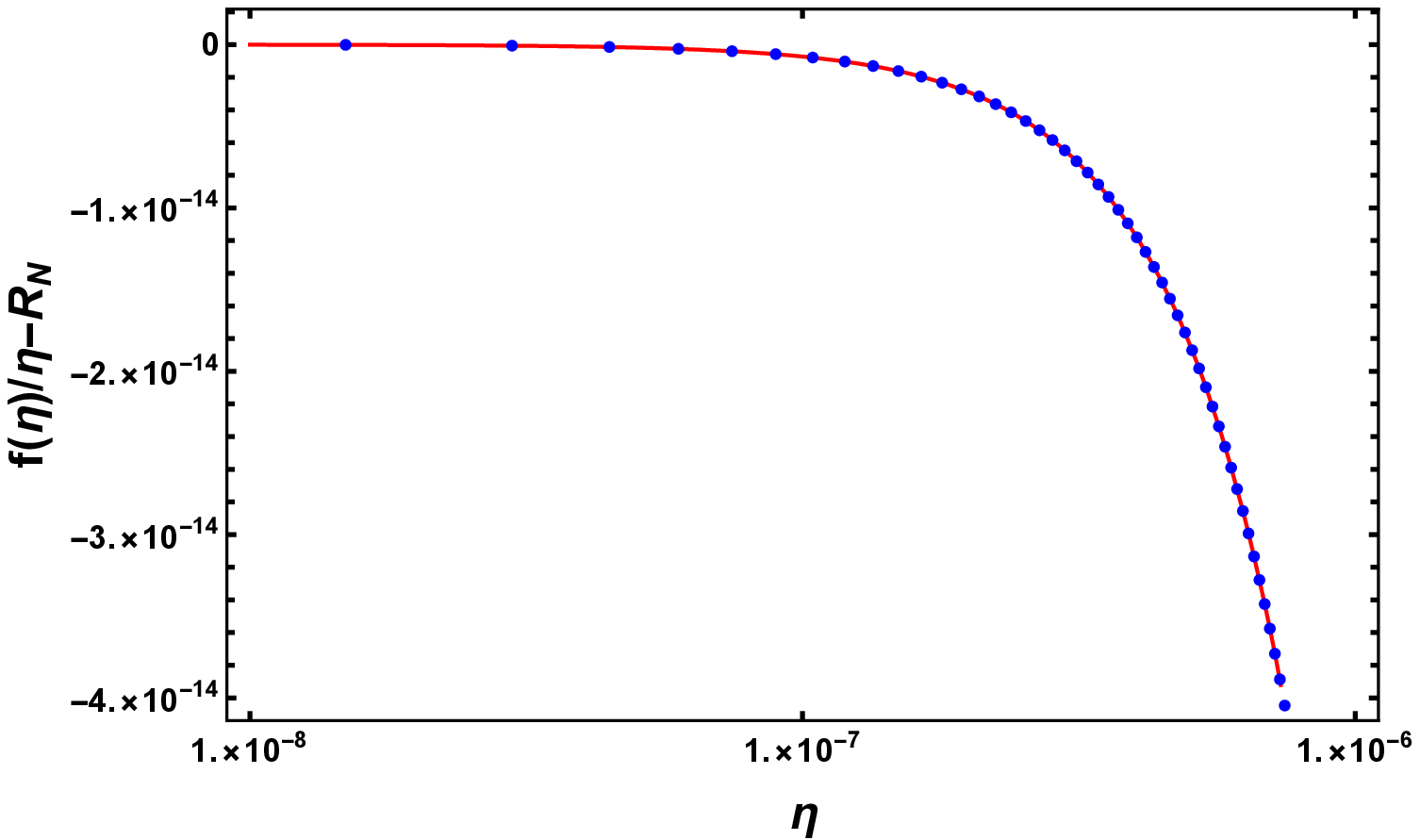}
\includegraphics[scale=0.375]{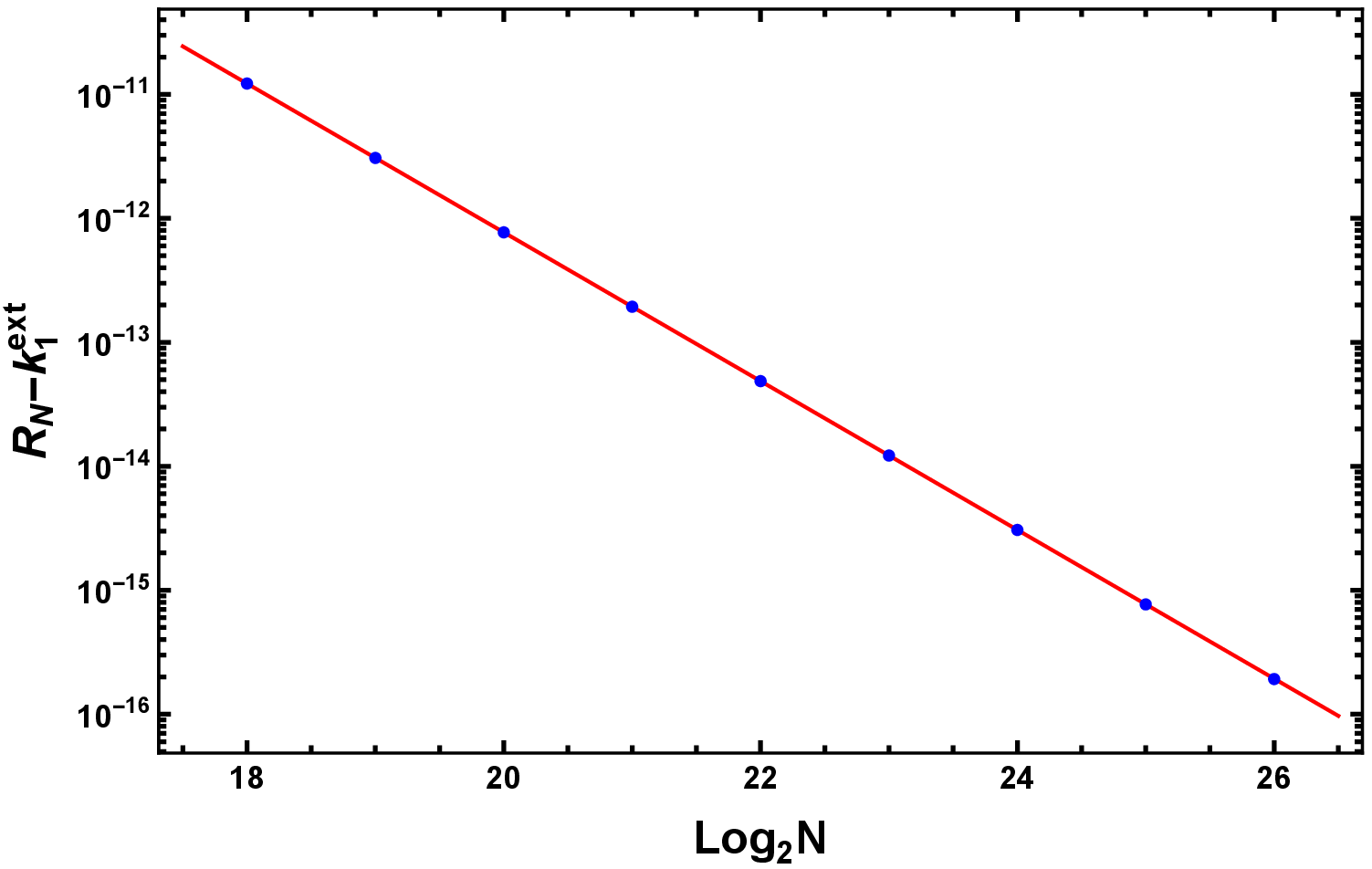}
\caption{Left: numerical results for $f(\eta)/\eta -R_N$ with $N=2^{26}$. Right: numerical results for $R_N-k_1^{\text{ext}} = \beta \gamma^{\log_2 N}$ as a function of $\log_2 N$.}
\label{fig:Num-ODE}
\end{figure}

The most important input for the approximate analytical solution is the value of connecting parameter $k_s$. We can get it from the numerical solution in the following way.
To determine $k_s=\lim_{\eta\rightarrow 0}f(\eta)/\eta^s$ with a higher precision,
the behavior of $f(\eta)$ near $\eta =0$ can be parameterized by
\bea
f(\eta) \rightarrow \left(\frac{\eta}{\sqrt{1+\eta^2}} + \tilde{h}(\eta)\right)\tanh^{s-1}(\eta),
\eea
For example, for winding number $s=1$, we find that $f(\eta)/\eta$ can be best fitted by $R_N(1-\eta^2/8)$ for small $\eta$, where $R_N$ is the fitting parameter corresponding to the solution with $N$ grid points. The left panel in FIG. \ref{fig:Num-ODE} shows the numerical result for $f(\eta)/\eta-R_N$ with $N=2^{26}$. The dependence of $R_N$ on $N$ is in Table \ref{rn-k1} and visualized in the right panel of FIG. \ref{fig:Num-ODE}.

\begin{table}
\begin{tabular}{|c|l|}
\hline
$N=2^n$ & \hspace{1cm} $R_N$ \\
\hline
n=18      & 0.583189495872 \\
n=19      & 0.5831894958634 \\
n=20      & 0.5831894958611 \\
n=21      & 0.58318949586052 \\
n=22      & 0.58318949586037 \\
n=23      & 0.583189495860341 \\
n=24      & 0.583189495860332 \\
n=25      & 0.5831894958603300 \\
n=26      & 0.58318949586032947 \\
$k_1^{\text{ext}}$    & 0.58318949586032928 \\
\hline
\end{tabular}
\caption{The numerical values of $R_N$ and $k_1^{\text{ext}}$ extrapolated from $R_N$.}
\label{rn-k1}
\end{table}

We may extract the value $k_1=\lim_{N\rightarrow \infty} R_N$ by an extrapolation because these points can be best fitted in the form $R_N-k_1^{\text{ext}} = \beta \gamma^{\log_2 N}$ with fitting parameters $k_1^{\text{ext}}$,$\beta$ and $\gamma$. Then $k_1^{\text{ext}}$ can be regarded as a high-precision approximation to the exact value of $k_1$. Its reliability will be discussed in following sections.

In the same way, the extrapolated values of $k_s$ for $s=2,3$ can be obtained
\bea
k_2^{\text{ext}}&=& 0.153099102859539,  \\
k_3^{\text{ext}}&=& 0.02618342072162.    
\eea
The precision of the connecting parameters is reduced for larger $s$, because it is harder to extract the small $\eta$ behavior of $f(\eta) \propto \eta^s$.
But in any way, the present results are precise enough for our purpose. We note that these connecting parameters can be calculated in a semi-analytical scheme \cite{Boisseau:2006dq}. A similar method is discussed in Sec.\ref{Sec:error}, in which the connecting parameters are traced by choosing the proper root of a polynomial with increasing order.  The biggest problem of such semi-analytical schemes is to find the right root of a high-order polynomial, which is computationally expensive, meanwhile the improvement on the precision of $k_s$ is very limited in comparison with the results from the extrapolation method.


\section{Two-point Pad$\acute{\text{E}}$ approximants for the vortex solution}
\label{Sec:2-point-pade}
Analytically, we may try to mimic the solution to Eq. \eqref{eq:vortex} in terms of Taylor expansion series from both points $\eta=0$ and $\eta=\infty$ simultaneously. This is the basic idea of two-point Pad$\acute{\text{e}}$ approximants. Without loss of generality,  we may define $g(\eta)$ by taking a shift 1/2 in $f(\eta)$,  $g(\eta)=f(\eta)-1/2$, and then expand $g(\eta)$ at $\eta=0$ and $\eta=\infty$ in the form
\bea
g(\eta) & = & \frac{c_0}{2} + \sum_{l=1}^{\infty} c_l \eta^l,  ~~~~~~~~   \eta \rightarrow 0,  \label{pade-zero} \\
g(\eta) & = & -\frac{c_0}{2} -\sum_{l=1}^{\infty} c_{-l} \eta^{-l},  ~~~   \eta \rightarrow \infty, \label{pade-infty}
\eea
where $c_0=-1$ and other coefficients $c_{-i},\cdots, c_i$ can be derived from Eq. \eqref{eq:vortex}. For $\eta=0$, by inserting Eq. \eqref{pade-zero} into Eq.\eqref{eq:vortex}, we find $c_{l}=0$ for $0< l < s$ and a recursive relation for $c_{l}$ for $l \ge s$,
\bea
c_{l+2}\left[(l+2)^2-s^2\right] + c_l - \sum^{k_1+k_2+k_3 =l}_{k_1,k_2,k_3\ge s} c_{k_1} c_{k_2} c_{k_3}=0, ~~~\text{for}~~~l\ge s.   \label{ci-positive}
\eea
The last term in the left-hand-side of Eq. \eqref{ci-positive} comes from the nonlinearity of Eq. \eqref{eq:vortex}. Without this term, the coefficients are just those of Bessel's functions of the first kind. For the complete form, we can read out that only the coefficients with even/odd indices can be nonzero for the even/odd $s$,
and their values are solely determined by the first nonzero one, which is $c_s=k_s$. Take $s=1$ as an example, we have
\bea
c_3 &=& -\frac{1}{8} c_1, \nonumber \\
c_5 &=& \frac{1}{192} (c_1+ 8 c_1^3),  \nonumber \\
c_7 &=& -\frac{1}{9216} (c_1+80c_1^3), \nonumber \\
c_9 &=& \frac{1}{737280} (c_1+656c_1^3+1152c_1^5).
\eea


For $\eta=\infty$, we insert Eq. \eqref{pade-infty} into Eq. \eqref{eq:vortex} and obtain $c_{-1}=0$  and a recursive relation for $c_{-l}$ with $l \ge 2$,
\bea
c_{-l+2}\left[(l-2)^2-s^2\right] + c_{-l} - \sum^{k_1+k_2+k_3 =l}_{k_1,k_2,k_3 \ge 0} c_{-k_1} c_{-k_2} c_{-k_3}=0, ~~~\text{for}~~~l\ge 2. \label{ci-negative}
\eea
It can be read out that all coefficients with odd indices are vanishing, the values of coefficients with even indices are solely determined by $s$. For the case $s=1$, we have
\be
c_{-2}=\frac{1}{2}, ~c_{-4}=\frac{9}{8},~c_{-6}=\frac{161}{16},~c_{-8}=\frac{24661}{128}.
\ee

\begin{figure}[ht] 
\centering
\includegraphics[scale=0.4]{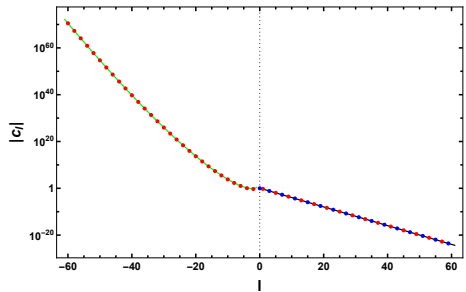}
\caption{The coefficients $c_l$ for $s=1$. The absolute values of $c_l$ are shown in a logarithmic scale, the sign of $c_l$ is indicated by the color of the point: red/blue stands for the positive/negative value. The green solid line stands for $1.92\times 2^{l/2}\Gamma[-l-1/2]$. The black solid line stands for $1.125\times 2.5121034^{-l}$. }
\label{fig:allci}
\end{figure}


In FIG. \ref{fig:allci}, we show the values of coefficients $c_l$ for $s=1$, where $c_1$ takes the value of $k_1^{\text{ext}}$. The absolute values of nonzero coefficients are shown in a logarithmic scale, the sign of each coefficient is indicated by the color of the point: red/blue stands for the positive/negative value. The magnitude of nonzero $c_l$ with $l\le0$ is fitted by $1.92\times 2^{l/2}\Gamma[-l-1/2]$ as the green solid line. The absolute values of nonzero $c_l$ with $l>0$ is fitted by  $1.125\times 2.5121034^{-l}$ as the black solid line. So that we can read out the convergence radius for the series in Eq. \eqref{pade-zero} is about $2.5$,  while the convergence radius for Eq. \eqref{pade-infty} is zero if we take $1/\eta$ as the expansion variable. It seems that polynomials of Taylor expansion cannot approximate the solution to Eq. \eqref{eq:vortex}. We have to seek a new type of function to incorporate the boundary conditions from both sides and describe the numerical solution in the whole range $0< \eta < +\infty$.


The two-point Pad$\acute{\text{e}}$ approximants in terms of rational functions have the form
\bea
\hat{g}_{i,j}(\eta) = \frac{P_{i,j}(\eta)}{Q_{i,j}(\eta)} = \frac{\sum_{l=0}^m \alpha_l \eta^l }{\sum_{l=0}^m \beta_l \eta^l}, ~~~\text{with}~~ m= \frac12(i+j).  \label{pade-2p}
\eea
In this form, $P_{i,j}(\eta)$ and $Q_{i,j}(\eta)$ are polynomials with the truncation order $m$, and $\alpha_l$ and $\beta_l$ are the coefficients determined by requiring that the expansion of $\hat{g}_{i,j}(\eta)$ at $\eta \rightarrow 0^+$ and $1/\eta \rightarrow 0^+$ agrees with Eq. (\ref{pade-zero}) and (\ref{pade-infty}) up to and including $c_{i-1}\eta^{i-1}$ and $c_{-j} \eta^{-j}$ respectively. So the residues of the approximants are of the order $i$ and $j+1$ at the boundaries, i.e.
\bea
\hat{g}_{i,j}(\eta)- g(\eta)  = {\cal{O}}(\eta^i, \eta^{-j-1}).
\eea
With $c_l$ in Eqs. (\ref{pade-zero},\ref{pade-infty}), the polynomials in the numerator and denominator in Eq. \eqref{pade-2p} can be constructed in a determinant representation \cite{Sidi:1980},
\bea
Q_{i,j}(\eta) = \left|
\begin{array}{ccc cccc}
1 & \eta & \cdots &  \eta^m \\
c_{i-1} & c_{i-2} &\cdots & c_{i-m-1} \\
c_{i-2} & c_{i-3} &\cdots & c_{i-m-2} \\
 \vdots   & \vdots   & \ddots & \vdots \\
c_{i-m}   & c_{i-m-1} &  \cdots  & c_{-j} \\
\end{array}
\right|,
\eea
and $P_{i,j}(\eta)$ is obtained from $Q_{i,j}(\eta)$ by replacing the first row with the vector
\bea
\left\{S_{m-1}(\eta), ~ \eta S_{m-2}(\eta), ~ \cdots, ~ \eta^{m-1}S_0(\eta), \eta^m T_{0}(\eta)\right\}  ~~~ &&\text{for} ~~~ i\ge m,  \\
\left\{T_0(\eta), ~ \eta T_1(\eta), ~ \cdots, ~\eta^{m-1}T_{m-1}(\eta),\eta^m T_{m}(\eta)\right\}  ~~~ &&\text{for} ~~~ i\le m,
\eea
where $S_k(\eta)$ and $T_k(\eta)$ are defined as
\bea
S_k(\eta)   &=& \frac{c_0}{2} + \sum_{l=1}^k c_l \eta^l,   \label{sk}\\
T_{k}(\eta) &=& -\frac{c_0}{2} - \sum_{l=1}^{k} c_{-l} \eta^{-l}. \label{tk}
\eea
Note that Eq. (\ref{sk}) and (\ref{tk}) are just the series of Eq. (\ref{pade-zero}) and (\ref{pade-infty}) up to $k$ respectively.

\begin{figure}[ht] 
\centering
\includegraphics[scale=0.32]{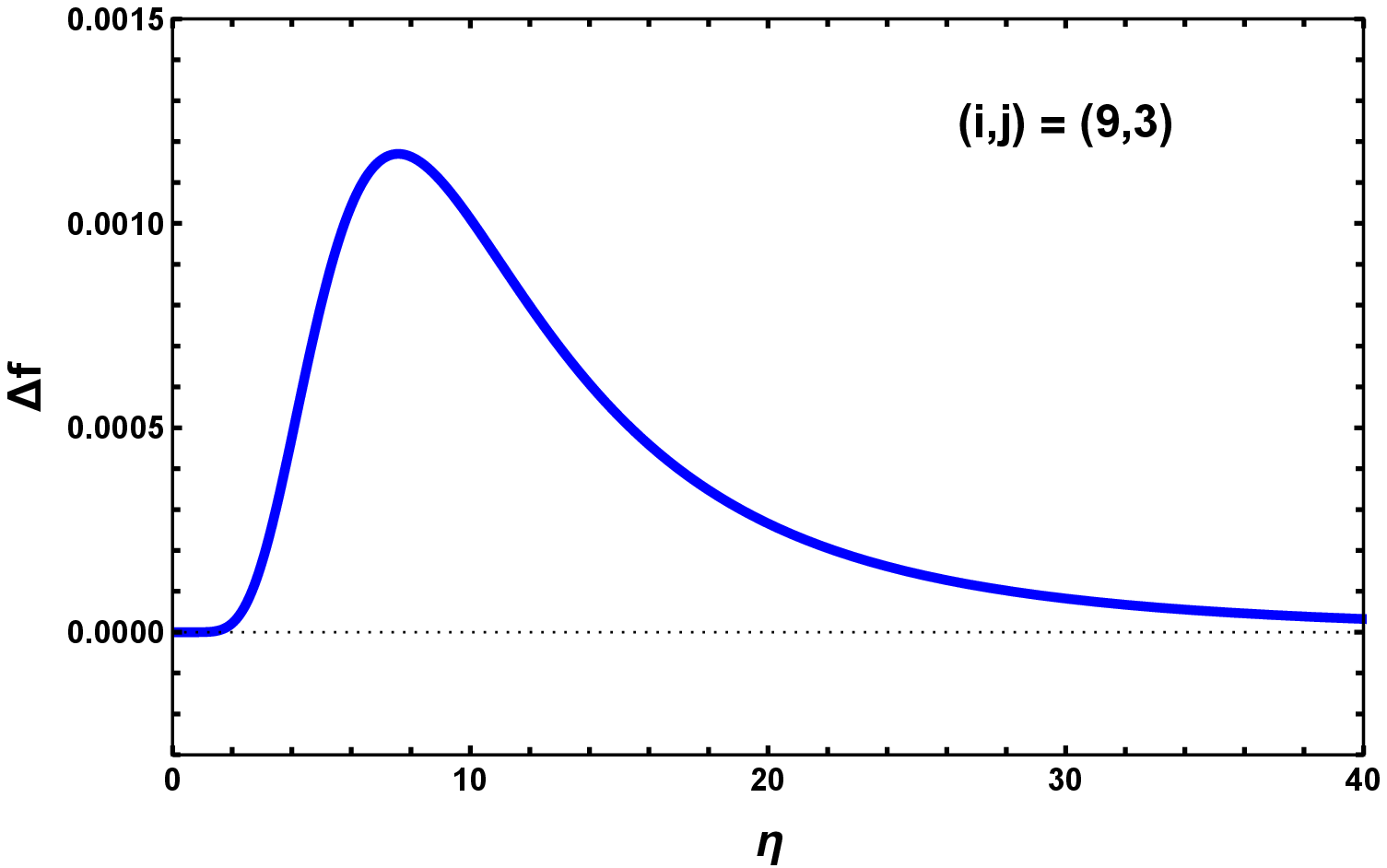}
\includegraphics[scale=0.32]{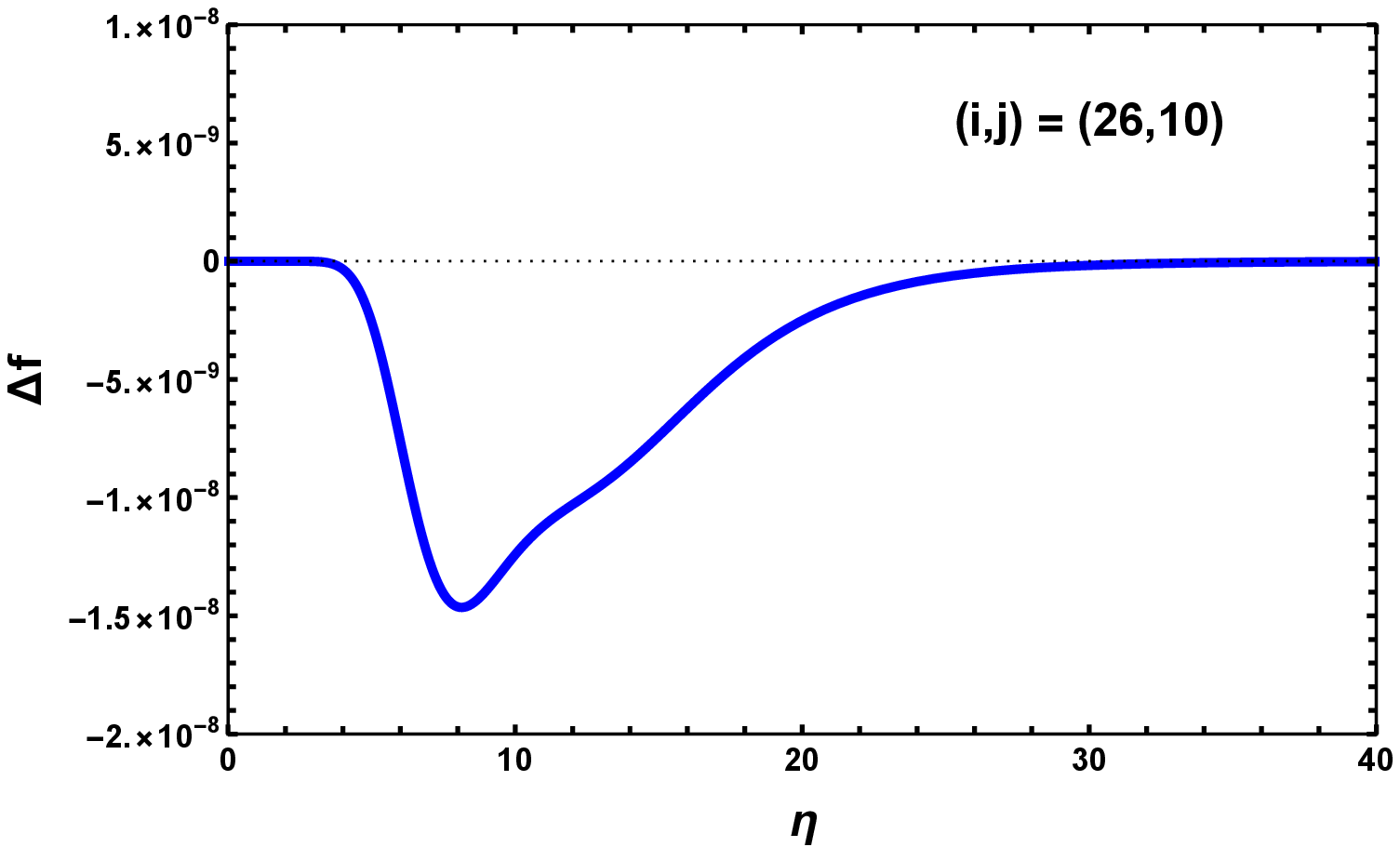}
\includegraphics[scale=0.32]{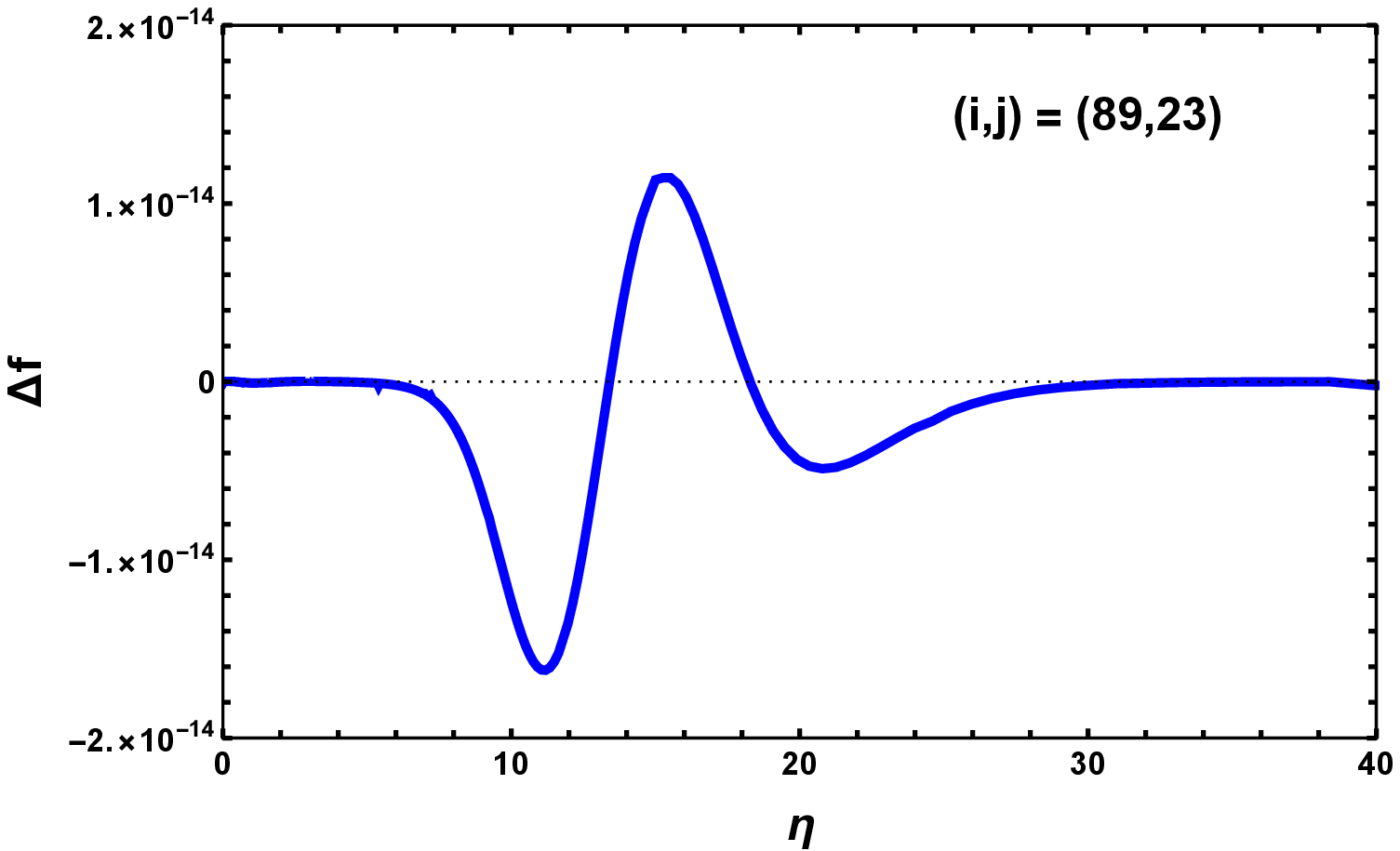}
\caption{The difference between the approximate functions and the "exact" solution for three typical truncation indices.}
\label{fig:diff-funcs}
\end{figure}

In such a way, the two-point Pad$\acute{\text{e}}$ approximants for the solution to the ODE \eqref{eq:vortex} can be easily constructed. The truncation index $(i,j)$ sets the size and structure of $P_{i,j}$ and $Q_{i,j}$, while the numerical value of each matrix element can be determined by the recursive relations in Eqs. (\ref{ci-positive}) and (\ref{ci-negative}) with the connecting parameter $c_s=k_s$. But in practice, the expected precision in comparison with the numerical solution is not guaranteed by the constructed function \eqref{pade-2p}, although $k_s$ can be determined with an extraordinary precision. The quality of the result depends on the choice of $(i,j)$, so we have to check whether the required accuracy can be achieved for particular values of $i$ and $j$. We regard the high precision numerical solution discussed in the previous section as an "exact" solution, from which the accuracy of the approximants can be estimated. For the primary vortex with $s=1$, an economic approximate function is given by setting $i=9$ and $j=3$, which is the best choice for $m = (i+j)/2 \le 6 $. The constructed function is parametrized with 12 coefficients in the form
\bea
\tilde{f}(\eta)=\frac{0.000536403 ~\eta ^6+0.00499313 ~\eta ^5+0.0279403 ~\eta ^4+0.109179 ~\eta
   ^3+0.297503 ~\eta ^2+0.583189 ~\eta }{0.000536403 ~\eta ^6+0.00499313 ~\eta
   ^5+0.0282085 ~\eta ^4+0.111676 ~\eta ^3+0.312211 ~\eta ^2+0.510131 ~\eta
   +1}.  \label{Pade-6}
\eea
The Taylor expansion of this fractional polynomial can reproduce $c_1, \cdots, c_{8}$ for $\eta\rightarrow 0$ and $c_{-1},\cdots,c_{-3}$ for $\frac{1}{\eta}\rightarrow 0$. The accuracy of the function with $i=9$ and $j=3$, defined by $\max[|\tilde{f}(\eta)-f(\eta)|]$, is about $1.2\times 10^{-3}$, as shown in the left panel of FIG. \ref{fig:diff-funcs}. It is much better than one general-purpose interpolation routine with limited input data points. The accuracy can be significantly improved for larger $m$, as shown in the middle panel of FIG. \ref{fig:diff-funcs}, the accuracy is about $1.5\times 10^{-8}$ for the function with $i=26$ and $j=10$. The coefficients in this function are listed in Appendix \ref{sec-data}. This approximate function with 36 parameters is good enough for the precise simulation of vortices within GPE.


In the same way, the two-point Pad$\acute{\text{e}}$ approximants for the profile functions of vortices with larger winding numbers can be achieved. The coefficients of the fractional polynomials for $s=2,3$ are given in Appendix \ref{sec-data}. In both cases, the accuracy is better than $10^{-6}$ in the whole range of $\eta$. It is interesting to see that the approximants with even $s$ contain only even terms in $P_{i,j}$ and $Q_{i,j}$ because the coefficients $c_l$ with odd $l$ are all vanishing in Eqs. (\ref{pade-zero}) and (\ref{pade-infty}). 


Before we work on even higher order functions, we must keep in mind that the coefficients will all change if we choose different $i$ and $j$. The reason is that the polynomials in the numerator or denominator result from matrix determinants, any change to the matrix will give very different polynomials. So we can not expect the coefficients of polynomials will be determined order by order as in a perturbation theory.


To explore the limit of this method, we try to calculate $P_{i,j}(\eta)$ and $Q_{i,j}(\eta)$ for large $i$ and $j$. 
The problem is that the matrices of $P_{i,j}(\eta)$ and $Q_{i,j}(\eta)$ involve coefficients $c_l$ that differ dramatically in magnitudes as shown in FIG. \ref{fig:allci}, it is a numerical challenge to keep the precision in calculating the determinant of such matrices of large dimension. We found Mathematica can handle these matrices with $i\le 100$ and $m\le 60$. The best approximate function that can be accessible corresponds to the matrix for $i=89$ and $j=23$, the accuracy is of the order $10^{-14}$, as shown in the right panel of FIG. \ref{fig:diff-funcs}.
The structure and coefficients of the rational functions may indicate some hints for the exact solution to the nonlinear ODE. In Appendix \ref{Sec:tanh}, we give an example on how the Pad$\acute{\text{e}}$ approximants help obtain an exact solution to a nonlinear ODE. But for Eq. \eqref{eq:vortex}, we fail to find any hint for an exact solution except approximate functions at very high precision.

\section{Error estimation and challenge}
\label{Sec:error}

The precision of Pad$\acute{\text{e}}$ approximants cannot be improved by simply increasing the truncation order. The asymptotic errors for both boundaries 
can be estimated as 
\bea
|g(\eta)-\hat{g}_{i,j}(\eta)| &=& \left|\frac{D_{i-m,m}}{D_{i-m-1,m-1}}\right|\eta^i(1+{\cal{O}}(\eta)) ~~~\text{for} ~~~ \eta \rightarrow 0, \label{D-1}  \\
|g(\eta)-\hat{g}_{i,j}(\eta)| &=& \left|\frac{D_{i-m-1,m}}{D_{i-m,m-1}}\right|\eta^{-j-1}(1+{\cal{O}}(\eta^{-1})) ~~~\text{for} ~~~ \eta \rightarrow \infty, \label{D-2}
\eea
with $D_{r,t}$ being the determinant of the coefficient matrix
\bea
D_{r,t} =\left|
\begin{array}{ccc cccc}
c_r & c_{r+1} & \cdots &  c_{r+t} \\
c_{r-1} & c_{r} &\cdots & c_{r+t-1} \\
 \vdots   & \vdots   & \ddots & \vdots \\
c_{r-t}   & c_{r-t+1} &  \cdots  & c_{r} \\
\end{array}
\right|.  \label{Eq:Drt}
\eea

\begin{figure}[ht] 
\centering
\includegraphics[scale=0.35]{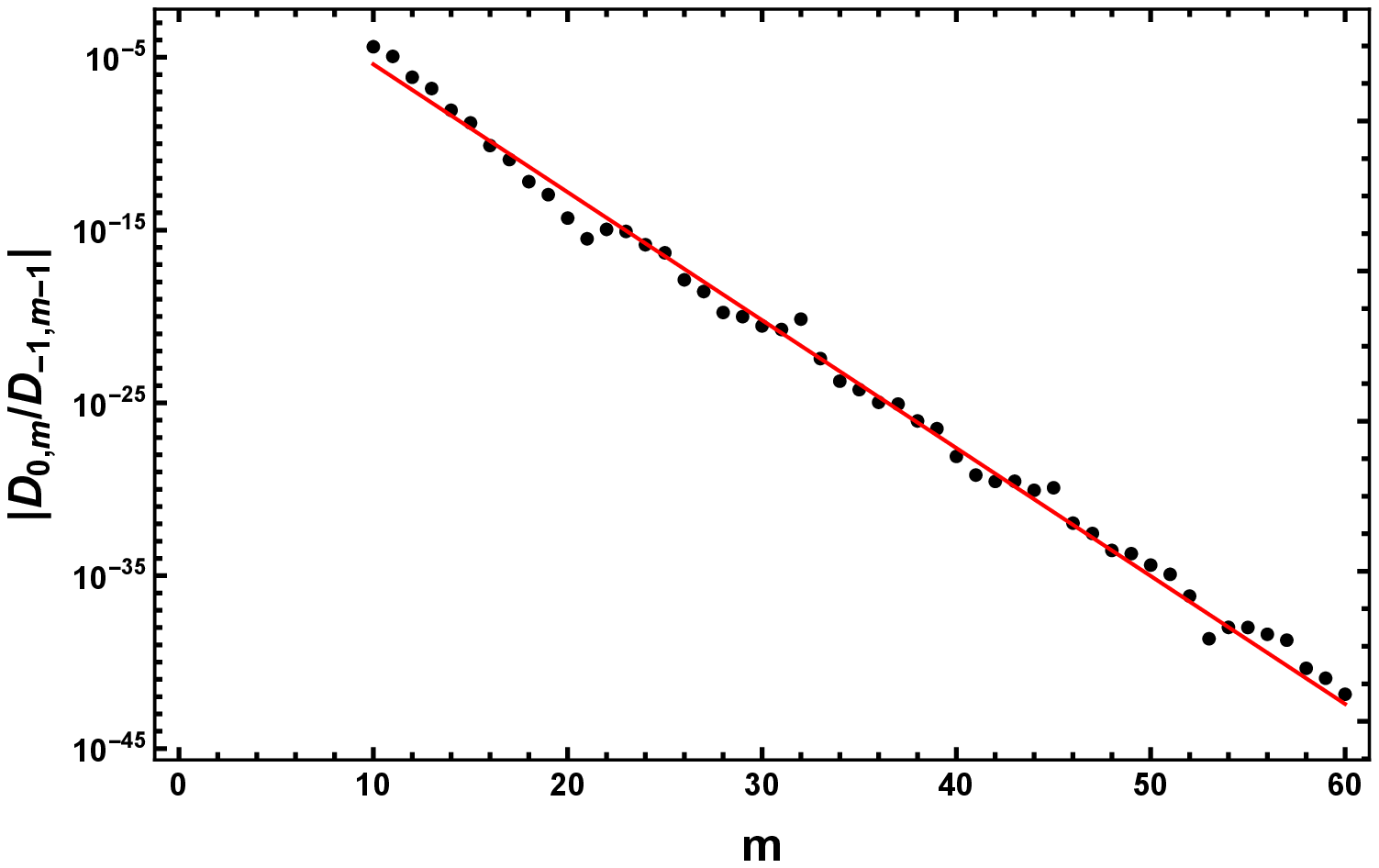}
\includegraphics[scale=0.35]{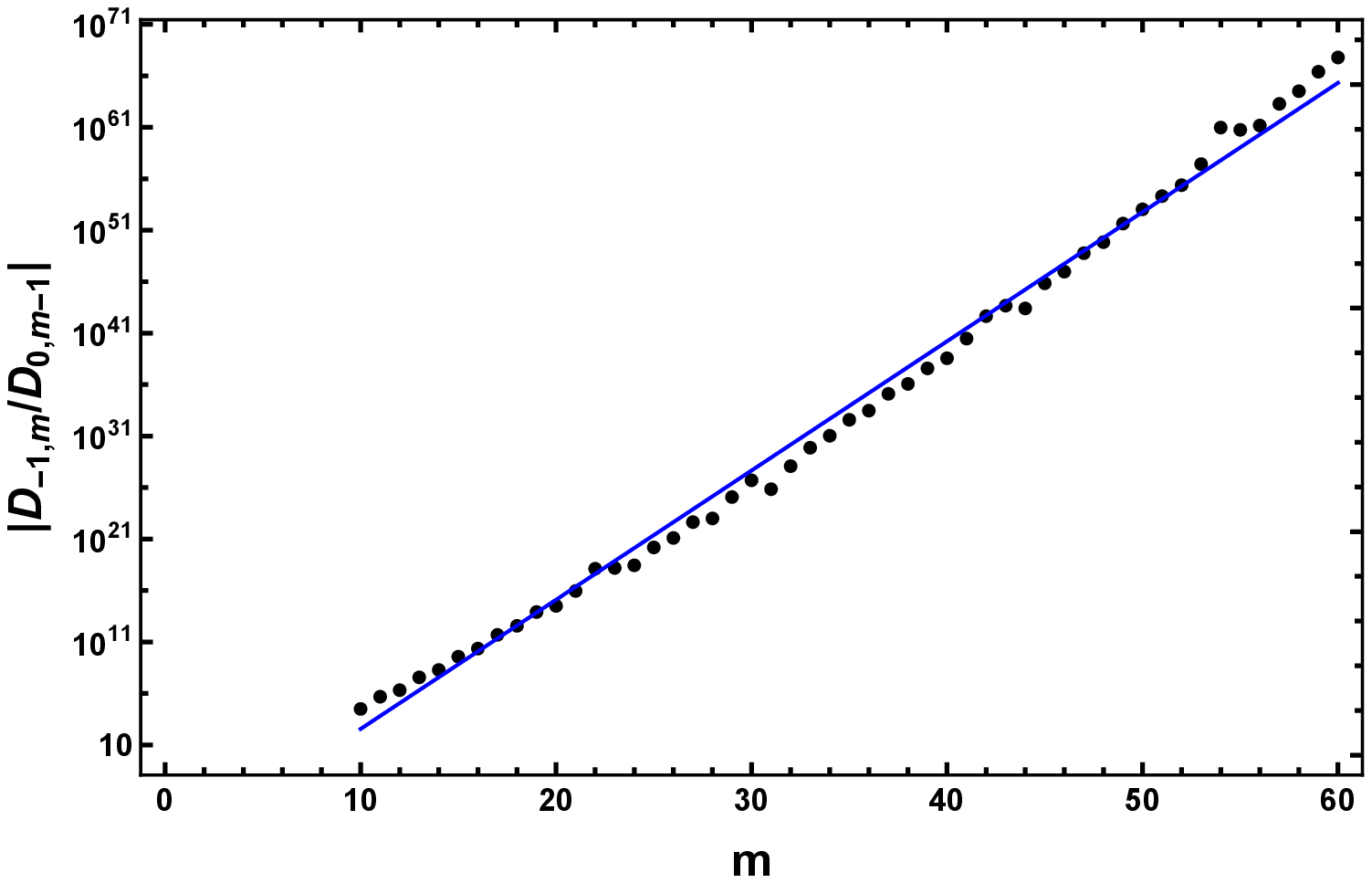}
\caption{Left: the ratio $|D_{0, m}/D_{-1,m-1}|$ (points) varies with $m$, the red line shows the function $10^2\times 5.5^{-m}$.  Right: the ratio $|D_{-1,m}/D_{0,m-1}|$ (points) varies with $m$, the blue line shows the function $10^{-10}\times 18^m$.}
\label{fig:Ratio-D}
\end{figure}

The error for small $\eta$ is controlled by the ratio $|D_{i-m, m}/D_{i-m-1,m-1}|$. To have an intuitive idea about the ratio, we estimate it as a function of $m$ for the case $i=j=m$. As shown in the left panel of FIG. \ref{fig:Ratio-D}, the ratio can be approximated as $10^2\times 5.5^{-m}$. This feature reveals that the convergence radius of $\hat{g}_{i,j}$ has been extended from $2.5$ to $5.5$ by Pad$\acute{\text{e}}$ approximants for small $\eta$. For large $\eta$, the dependence of $|D_{-1,m}/D_{0,m-1}|$ on $m$ is shown in the right panel of FIG. \ref{fig:Ratio-D} and can be described by $10^{-10}\times 18^m$. This means that the error is under control for $\eta > 18$, or the convergence radius for $1/\eta$ is extended from 0 to 1/18. Although the change of the convergence radius is small, the coverage range of the approximants is expanded significantly. Actually, this is a remarkable progress of Pad$\acute{\text{e}}$ approximants which aim to reproduce a complicated function from both boundaries simultaneously. Inside the gap between two convergence regions, the rational function connects both regions smoothly, which should reproduce the target function with a similar or slightly worse precision if the target function does not contain a singularity in the gap. The error can be further reduced by fine-tuning of $i$ and $j$ if the target function is known with a higher precision. As we have already shown in FIG. \ref{fig:diff-funcs}, the vortex profile function can be reproduced with a high accuracy, and the behaviors of errors can be well understood as discussed above.

\begin{figure}[ht] 
\centering
\includegraphics[scale=0.4]{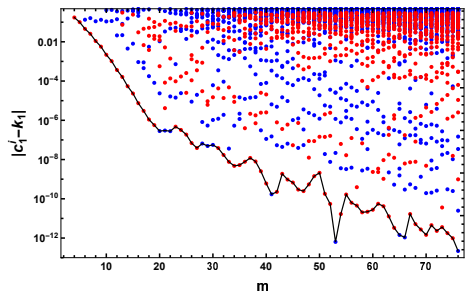}
\caption{The relative distance between $k_1$ and the roots of $D_{0,m}$ for different $m$. The magnitude is shown in logarithm scale, while the the sign is indicated by color: red points stand for $c_1^i > k_1$, blue points stand for $c_1^i < k_1$. The bottom line indicates the chain of the closest roots, which is technically traceable without knowing $k_1$ at first. }
\label{fig:diff-roots}
\end{figure}

\begin{figure}[ht]
\centering
\includegraphics[scale=0.4]{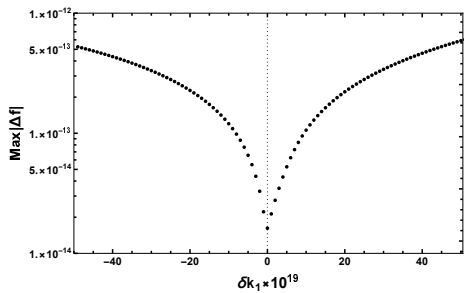}
\caption{The maximum deviation of the approximants from the "exact" numerical solution as a function of the variation of $k_1$.}
\label{fig:min-scan-k1}
\end{figure}

In the description of Pad$\acute{\text{e}}$ approximants for vortex profiles, we find that $P_{i,j}$, $Q_{i,j}$ and $D_{r,t}$ are sensitive to the connecting parameter $k_s$. Since the corresponding matrices have similar structure, we take $D_{r,t}$ as an example to discuss its dependence on $k_s$. For simplicity, we look at $D_{0,m}$ and $k_1$. From Eq.  \eqref{Eq:Drt}, $D_{0,m}$ can be expressed as a polynomial of $c_1$ and the order increases fast with the dimension of the matrix. We find that $D_{0,m}$ has some real roots, 
denoted as $c_1^i$, which are very close to $k_1$. The distance between each $c_1^i$ and $k_1$ is shown in FIG. \ref{fig:diff-roots}, where the magnitude is plotted in logarithm scale and the sign is marked by color. A bunch of roots near $k_1$ indicate that $D_{0,m}$ varies dramatically when $c_1$ approaches $k_1$. So it can be understood as that a small deviation from $k_1$ will lead to an uncontrollable variation of $D_{r,t}$ when the dimension of the matrix becomes large. Similar behavior is also observed in $P_{i,j}$ and $Q_{i,j}$ as functions of $c_1$. So the accuracy of the Pad$\acute{\text{e}}$ approximants for large $m$ strongly depends on the precision of $k_1$. As shown in FIG. \ref{fig:min-scan-k1}, for the best result that we find (the right panel of FIG. \ref{fig:diff-funcs}), a very small change in the value of $k_1$ leads to a significant increase in the error of the approximants. So the precision of $k_1$ is crucial for that of Pad$\acute{\text{e}}$ approximants for large $m$. 


The minimal distance in FIG. \ref{fig:diff-roots} can be extremely small for large $m$, it seems that this feature can be employed to find the asymptotic value of $k_1$ from $c_1^i$ with a carefully chosen initial value and iteration of increasing $m$. Technically, we can set the only possible root of $D_{0,4}$ as the initial value, and replace it with one root of $D_{0,5}$ which is real and closest to the original one. Repeat this procedure, a chain of the closest root can be found, approaching the true value of $k_1$. As shown in FIG. \ref{fig:diff-roots}, the bottom line is traceable without knowing $k_1$ at first. This idea is workable for $m<80$, with the precision at the level of $10^{-12}$. But for even larger $m$, it is difficult to find the right root of a high-order polynomial, because the density of roots near $k_1$ increases while the minimal distance between them decreases exponentially, which raises the risk of finding wrong roots. This is the reason why we do not seek the value of $k_1$ from roots of $D_{r,t}$ but import the extrapolated value from a series of numerical solutions with a reliable precision. At present stage, with Mathematica running on a laptop, we can confirm the value of $k_1$ with 17 significant digits, which helps us to promote the precision of the two-point Pad$\acute{\text{e}}$ approximants to $10^{-14}$. The method may help us to search for the exact solution of the vortex profile. 


\section{summary and conclusion}
\label{Sec:summary}
Understanding the vortex dynamics is important for the study of superfluids and many other physical systems. Due to the nonlinear nature of the vortex dynamics, the evolution of vortex can only be studied in a perturbative way, which requires a high precision solution to the stationary wave function of the vortex. The nonlinear ordinary differential equation for the vortex profile function can be solved numerically with high precision.

For practical applications, we have constructed semi-analytical solutions to this problem with two-point Pad$\acute{\text{e}}$ approximants. These solutions are presented with simple rational functions, which have a high accuracy and are ready for use in the precise simulation of the vortex dynamics. The coefficients of rational functions strongly depend on the value of the connecting parameter which can be obtained in a high precision from an extrapolation of the numerical solution or a systematic root-finding in a series of polynomials.

The accuracy of the approximate solutions can be improved significantly with a reasonable extension in the order of rational functions.  With a systematic scan of the truncation orders, the best accuracy is found to be at the order of $10^{-14}$. The errors of the approximate functions and the limitation of two-point Pad$\acute{\text{e}}$ approximants are discussed, which can extend our understanding on the nonlinearity of the vortex dynamics. This investigation may provide the clue for an exact solution to the vortex profile function.


The methods and algorithms developed in this work for high precision solutions to the vortex equation can be applied to other nonlinear systems, such as the Schrödinger-Poisson problem \cite{Braaten:2015eeu,Levkov:2018kau}, the shape and properties of Bose stars formed by dark matter through the universal gravitational interaction and so on.

\paragraph*{Acknowledgments: }
J.D.\ is supported by the Natural Science Foundation of Shandong Province under Grant No.\ ZR2020MA099. Q.W.\ is supported in part by the National Natural Science Foundation of China (NSFC) under Grants No.\ 12135011, 11890713 (a subgrant of 11890710), and by the Strategic Priority Research Program of the Chinese Academy of Sciences (CAS) under Grant No.\ XDB34030102.

\appendix

\section{An example for exact solution with two-point Pade approximants}
\label{Sec:tanh}
We consider the nonlinear ODE
\bea
\frac{\text{d}^2}{\text{d}x^2}f(x) +f(x) - f^3(x) =0, \label{ODE-tanh}
\eea
with the boundary conditions $f(0)=0$ and $f(\infty)=1$. At the zero point, we may insert the formal solution
\be
f(x)=\sum_0^{\infty}c_i x^i ,
\ee
into the ODE. It is easy to check that all even terms are vanishing, i.e. $c_{2k}=0$ for $k=0,1, \cdots $.
The odd terms satisfy the recurrence relation
\bea
c_{n+2}(n+2)(n+1) + c_n - \sum_{k_1+k_2+k_3 =n} c_{k_1} c_{k_2} c_{k_3}=0.
\eea
The only unsettled parameter is $c_1$, by which other coefficients can be expressed, for example, we have the following expressions for $c_3, \cdots , c_9$ in terms of $c_1$
\bea
c_3 &=& -\frac{1}{3!} c_1 ,\nonumber \\
c_5 &=& \frac{1}{5!} (c_1+6 c_1^3) ,  \nonumber \\
c_7 &=& -\frac{1}{7!} (c_1+66c_1^3) ,\nonumber \\
c_9 &=& \frac{1}{9!} (c_1+612c_1^3+756c_1^5) .
\eea
We may represent the solution in the form of Pad$\acute{\text{e}}$ approximants
\bea
\tilde{f}^{[N]}(x)=\frac{\sum_{i=0}^N a_i x^i}{\sum_{j=0}^{N} b_j x^j}. \label{Pade-form1}
\eea
Because the solution is determined by the ratio, it is natural to demand $b_0=1$.
Because $c_0=0$, we can set $a_0=0$. Other coefficients should satisfy
\bea
a_i = \sum_{j=0}^{N} b_j c_{i-j}, ~~~~~\text{for}~~  1 \le i \le N.  \label{eq:ab1}
\eea

For the boundary condition at $x\rightarrow \infty$, we can change the variable to $z=1/x$, so that Eq. \eqref{ODE-tanh} can be rewritten as
\bea
z^4 \frac{\text{d}^2}{\text{d}z^2}f(z) +2z^3 \frac{\text{d}}{\text{d}z}f(z)   +f(z)  - f^3(z) =0 .
\eea
Then we can take the formal solution in the form
\be
f(z)=\sum_0^{\infty}d_i z^i .
\ee
It is easy to obtain $d_0=1$ and $d_{i}=0$ for $i>0$.
So there is another constraint for coefficients
\bea
a_i = \sum_{j=0}^{N} b_j d_{j-i} = b_i, ~~~~~\text{for}~~  1 \le i \le N.  \label{eq:ab2}
\eea
Combining Eq.\eqref{eq:ab1} and Eq.\eqref{eq:ab2}, we obtain a set of linear equations for $(a_1,\cdots, a_N)$ or $(b_1,\cdots, b_N)$. We may write the equations into a matrix form. As an example, the equations for $N=5$ read
\bea
\left(
\begin{array}{ccccccccc}
 1  & 0 & 0 & 0 & 0  \\
 -c_1 & 1  & 0 & 0 & 0    \\
 0 & -c_1 & 1  & 0 & 0   \\
 \frac{c_1}{6} & 0 & -c_1 & 1  & 0 \\
 0 & \frac{c_1}{6} & 0 & -c_1 & 1
\end{array}
\right)
\left(
\begin{array}{c}
 b_1 \\
 b_2 \\
 b_3 \\
 b_4 \\
 b_5
\end{array}
\right)
=\left(
\begin{array}{c}
c_1 \\
 0 \\
 -\frac{c_1}{6} \\
 0 \\
 \frac{c_1^3}{20}+\frac{c_1}{120}
\end{array}
\right).
\eea
Because the matrix is already in a triangle form, the solution can be easily obtained
\bea
\left(
\begin{array}{c}
 a_1 \\
 a_2 \\
 a_3 \\
 a_4 \\
 a_5
\end{array}
\right)
=\left(
\begin{array}{c}
 b_1 \\
 b_2 \\
 b_3 \\
 b_4 \\
 b_5
\end{array}
\right)
=\left(
\begin{array}{c}
 c_1 \\
 c_1^2 \\
 c_1^3-\frac16 c_1 \\
 c_1^4-\frac13 c_1^2 \\
 c_1^5-\frac{9}{20}c_1^3+\frac{1}{120}c_1
\end{array}
\right). \label{eq:a-b}
\eea
To determine the value of $c_1$, we may demand one more constraint on the solution by taking $a_{N+1}=b_{N+1}=0$, which can be written as
\bea
\sum_{j=1}^{N} b_j c_{N+1-j}+c_{N+1} =0.
\eea
It gives an algebra equation for $c_1$. As an example, for $N=5$, the equation reads
\bea
\frac{c_1^2}{90}\left(90c_1^4  -51 c_1^2 +4 \right)=0.
\eea
The meaningful root to the above equation is $\sqrt{(51+3\sqrt{123})/180}\approx 0.687481$.
In this way, we can get a series of $c_1$ values
$0.69976, 0.704416,0.706143,0.706768,0.706989$ for $N=6,7,8,9,10$ respectively.
We may check that the approximate value of $c_1$ becomes closer to $\sqrt{2}/2 \approx 0.707107$ for large $N$.
If we set $c_1=\sqrt{2}/2$, we find the solution in Eq. \eqref{eq:a-b} can be simplified to $a_i=b_i= (\sqrt{2})^i/(2\cdot i!)$. So the solution can be put into the form
\bea
\tilde{f}^{[N]}(x)=\frac{\sum_{i=1}^N (\sqrt{2}x)^i/(2\cdot i!)}{1+\sum_{i=1}^{N} (\sqrt{2}x)^i/(2\cdot i!)}
=\frac{\sum_{i=0}^N  (\sqrt{2}x)^i/i! -1}{\sum_{i=0}^N (\sqrt{2}x)^i/i! +1}.
\eea
It happens to give $f(x)=\tanh(x/\sqrt{2})$ if we send $N$ to $\infty$.


\section{Coefficients in approximate functions for vortex profiles} \label{sec-data}
In this appendix we list the coefficients in approximate functions in \eqref{pade-2p} with which one can reproduce the vortex profiles in high accuracy.
\bea
\begin{array}{|c|c|c|c|c|}
   \hline
   &  \multicolumn{2}{c|}{s=1}  & \multicolumn{2}{c|}{s=3} \\
   \hline
l  &  \alpha_l  & \beta_l & \alpha_l  & \beta_l \\
   \hline
0  &  0                           &  1.000000000                &  0                            &    1.000000000                  \\
1  &  5.831894959\times 10^{-1}   &  5.873264261\times 10^{-1}  &  0                            &    4.062542828\times 10^{-1}    \\
2  &  3.425226024\times 10^{-1}   &  5.455940613\times 10^{-1}  &  0                            &    2.100817734\times 10^{-1}    \\
3  &  2.452860386\times 10^{-1}   &  2.468318072\times 10^{-1}  &  2.618342072\times 10^{-2}    &    6.454678403\times 10^{-2}    \\
4  &  1.011343919\times 10^{-1}   &  1.145178304\times 10^{-1}  &  1.063712681\times 10^{-2}    &    2.069992436\times 10^{-2}    \\
5  &  3.831447420\times 10^{-2}   &  3.898275190\times 10^{-2}  &  3.864195663\times 10^{-3}    &    5.236196331\times 10^{-3}    \\
6  &  1.137856193\times 10^{-2}   &  1.179179362\times 10^{-2}  &  1.025235177\times 10^{-3}    &    1.313346436\times 10^{-3}    \\
7  &  2.909882040\times 10^{-3}   &  2.971398890\times 10^{-3}  &  2.391152072\times 10^{-4}    &    2.871142311\times 10^{-4}    \\
8  &  6.323721506\times 10^{-4}   &  6.423893504\times 10^{-4}  &  4.809356698\times 10^{-5}    &    5.537192270\times 10^{-5}    \\
9  &  1.180693171\times 10^{-4}   &  1.194567590\times 10^{-4}  &  8.539789317\times 10^{-6}    &    9.539350138\times 10^{-6}    \\
10 &  1.905344120\times 10^{-5}   &  1.922168115\times 10^{-5}  &  1.358658441\times 10^{-6}    &    1.482637465\times 10^{-6}    \\
11 &  2.672381372\times 10^{-6}   &  2.690240172\times 10^{-6}  &  1.934240689\times 10^{-7}    &    2.074012815\times 10^{-7}    \\
12 &  3.271123471\times 10^{-7}   &  3.287644644\times 10^{-7}  &  2.463814537\times 10^{-8}    &    2.607377411\times 10^{-8}    \\
13 &  3.497660717\times 10^{-8}   &  3.510863602\times 10^{-8}  &  2.842462008\times 10^{-9}    &    2.973581152\times 10^{-9}    \\
14 &  3.254331773\times 10^{-9}   &  3.263312251\times 10^{-9}  &  2.975340172\times 10^{-10}   &    3.082927325\times 10^{-10}   \\
15 &  2.613493838\times 10^{-10}  &  2.618418071\times 10^{-10} &  2.742356229\times 10^{-11}   &    2.830506106\times 10^{-11}   \\
16 &  1.786780526\times 10^{-11}  &  1.788474199\times 10^{-11} &  2.276667859\times 10^{-12}   &    2.335377575\times 10^{-12}   \\
17 &  9.848466014\times 10^{-13}  &  9.848466014\times 10^{-13} &  1.958886160\times 10^{-13}   &    1.958886160\times 10^{-13}   \\
18 &  3.387345177\times 10^{-14}  &  3.387345177\times 10^{-14} &  1.304660342\times 10^{-14}   &    1.304660342\times 10^{-14}   \\
\hline
\end{array} \nonumber
\eea

\bea
\begin{array}{|c|c|c|}
   \hline
   &  \multicolumn{2}{c|}{s=2}  \\
   \hline
l  &  \alpha_l  & \beta_l  \\
   \hline
0  & 0                          &  1.000000000                \\
2  & 1.530991029\times 10^{-1}  &   4.398930447\times 10^{-1} \\
4  & 5.458897193\times 10^{-2}  &   6.936791316\times 10^{-2} \\
6  & 5.406591644\times 10^{-3}  &   6.042107443\times 10^{-3} \\
8  & 2.685749628\times 10^{-4}  &   2.859576645\times 10^{-4} \\
10 & 7.837336242\times 10^{-6}  &   8.153179530\times 10^{-6} \\
12 & 1.472555945\times 10^{-7}  &   1.513096732\times 10^{-7} \\
14 & 1.934727229\times 10^{-9}  &   1.970129182\times 10^{-9} \\
16 & 1.693952735\times 10^{-11} &   1.724410709\times 10^{-11}\\
18 & 1.522898667\times 10^{-13} &   1.522898667\times 10^{-13}\\
   \hline
\end{array} \nonumber
\eea


\end{document}